\documentclass[12pt,a4paper]{article}
\usepackage{jheppub}
\title{Forced Fluid Dynamics from Gravity in Arbitrary Dimensions}
\author{T. Ashok}
\affiliation{Department of Applied Mathematics and Theoretical Physics, University of Cambridge, Cambridge CB3 0WA, England}
\emailAdd{A.Thillaisundaram@damtp.cam.ac.uk}

\abstract{
We consider long wavelength solutions to the Einstein-dilaton system with negative cosmological constant which are dual, under the AdS/CFT correspondence, to solutions of the conformal relativistic Navier-Stokes equations with a dilaton-dependent forcing term. Certain forced fluid flows are known to exhibit turbulence; holographic duals of forced fluid dynamics are therefore of particular interest as they may aid efforts towards an explicit model of holographic steady state turbulence. In recent work, Bhattacharyya \textit{et al} have constructed long wavelength asymptotically locally AdS$_5$ bulk spacetimes with a slowly varying boundary dilaton field which are dual to forced fluid flows on the $4-$dimensional boundary. In this paper, we generalise their work to arbitrary spacetime dimensions; we explicitly compute the dual bulk metric, the fluid dynamical stress tensor and Lagrangian to second order in a boundary derivative expansion.}

\begin{document}
\maketitle

\section{Introduction} 

The AdS/CFT correspondence \cite{Maldacena1,Maldacena2} offers a novel perspective on quantum field theories. This duality enables us to translate a problem in field theory into a gravitational/string theoretic language. The hope of course is that, where conventional field theoretic methods have failed, the dual description may prove to be more intuitive and the problem therefore more tractable. This is especially true for strongly coupled field theories and there is significant interest in obtaining holographic descriptions of certain strongly coupled phenomena \cite{Gubser1,Hartnoll1}.

Considerable progress has already been made in obtaining the holographic duals of equilibrium field theory configurations. Two canonical examples in the AdS/CFT dictionary are the Schwarzschild-AdS black brane which corresponds to a thermal state in the field theory and the Reissner-Nordstrom-AdS black brane which corresponds to a thermal state at finite density. Now, while an understanding of equilibrium states is useful to describe certain physical phenomena (the study of phase transitions, for example, need not include any explicit time-dependence), many interesting strongly coupled phenomena are dynamical. Hence, there is considerable physical motivation for the holographic study of non-equilibrium behaviour \cite{R1}.

Unfortunately, while the holographic study of non-equilibrium dynamics is of much greater interest, it is also correspondingly much more difficult. Some headway can be made by focusing on small deviations away from equilibrium. The study of these small amplitude perturbations is known as linear response theory and the holographic methodology involved is part of the standard AdS/CFT toolkit \cite{Kraus1, Witten1, Skenderis1, Skenderis2, Liu1, Liu2}. However, the regime of validity of linear response theory does not cover large amplitude, violent perturbations away from equilibrium and, in such cases, often progress can only be made using numerical methods \cite{Chesler1, Yaffe1, Yaffe2}.

If we are motivated by the desire to obtain \textit{analytically} the holographic dual of a certain class of interesting, non-trivial non-equilibrium phenomena (beyond the reach of linear response theory), a natural starting point would be fluid dynamics. We will now provide some intuition for this statement. For the sake of having a concrete example, we consider the most familiar case of the AdS/CFT correspondence: the duality between \textit{SU(N)} $\mathcal{N}=4$ Super Yang-Mills theory and Type IIB string theory on AdS$_5\times$S$^5$. For generic values of $N$ and coupling $\lambda$, both sides of this duality are fairly complicated theories. If we are interested in obtaining analytic, time-dependent solutions with the aim of studying non-equilibrium phenomena, it is well worth considering a limit in which the dynamics will simplify. A natural way forward would be to take $N \rightarrow \infty$ in the 't Hooft limit; the bulk theory now becomes classical Type IIB string theory. If we further take the strong coupling limit ($\lambda \rightarrow \infty$), the massive string states decouple and the bulk theory simplifies to Type IIB supergravity. Now, while progressing from Type IIB string theory on AdS$_5\times$S$^5$ to Type IIB supergravity certainly is a step in the right direction, more can still be done. Type IIB supergravity on AdS$_5\times$S$^5$ has several consistent truncations to reduced, decoupled subsectors of dynamics; we focus on the simplest of these which is pure Einstein gravity with negative cosmological constant\footnote{The bulk spacetime has $d+1$ dimensions. Also, we have set the AdS curvature radius to unity.},
\begin{equation} \label{Einstein1}
E_{AB} \equiv R_{AB} - \frac{1}{2}Rg_{AB} + \Lambda g_{AB} = 0, \;\;\;\;\Lambda \equiv -\frac{d(d-1)}{2}.
\end{equation}

It is worth emphasising here that this result applies with much greater universality than implied above. 
There are an infinite number of field theories possessing gravitational duals; all of which admit large $N$ and strong coupling limits. And in these limits, the bulk theories will generically simplify to two derivative Einstein gravity interacting with other fields. Regardless of the specific nature of these interactions, these bulk theories of gravity will certainly admit AdS$_{d+1} \times M_I$ as a solution ($M_I$ is some internal manifold). Bulk dynamics with these characteristics all possess consistent truncations to pure Einstein gravity with negative cosmological constant. In this sense, the dynamics described by equation (\ref{Einstein1}) is the universal subsector of dynamics for an infinite class of bulk theories. And from the field theory perspective, bearing in mind that the bulk graviton is dual to the boundary field theory stress tensor, we see that pure Einstein gravity with negative cosmological constant is the universal dual bulk description of the stress tensor dynamics of an infinite class of strongly coupled field theories. 

Let us now pause to summarise our current position. Motivated by our desire to obtain analytic, time-dependent holographic solutions, we were led to the universal subsector of dynamics (\ref{Einstein1}) which is the dual dynamics of the boundary stress tensor. Yet even now, attempting to classify all time-dependent solutions to the Einstein equations is far from an easy task. And on the field theory side as well, the full behaviour of the stress tensor is still very nontrivial. To achieve further progress it is again worth limiting our attention to a simpler case. A promising path that we could take would be to focus only on stress tensor dynamics for field theory configurations which are locally equilibriated. Such configurations are governed by fluid dynamics \cite{L}; and the key fluid dynamical equations of motion simply follow from conservation of the stress tensor\footnote{Greek indices label boundary coordinates.},

\begin{equation}
\nabla_\mu T^{\mu\nu}=0.
\end{equation}
Perhaps constructing bulk time-dependent solutions of (\ref{Einstein1}) dual to boundary fluid dynamics is a more realistic aim?

This goal was concretely achieved in \cite{B1} where the authors explicitly constructed asymptotically AdS$_5$ long wavelength solutions to the Einstein equations with negative cosmological constant which are dual to solutions of the four-dimensional conformal relativistic Navier-Stokes equations. It should be stressed here that this work constitutes a derivation of \textit{nonlinear} fluid dynamics from gravity and thus is valid for fluid dynamical solutions with arbitrarily large velocity amplitudes. This is distinct from previous work on holographic linearised hydrodynamics \cite{Son1,Son2,Son3,Son4} which is only valid for small amplitude perturbations about equilibrium configurations. Work on obtaining the holographic dual of nonlinear fluid dynamics was in some sense pioneered by \cite{Janik1,Janik2,Janik3}; here, the authors considered nonlinear solutions dual to Bjorken flow, a particular boost invariant flow.

This duality between long wavelength solutions of the Einstein equations and solutions of nonlinear boundary fluid dynamics has become known as the \textit{fluid/gravity correspondence} \cite{R2,R3}.
Subsequent work soon after the seminal paper \cite{B1} generalised this map to arbitrary spacetime dimensions \cite{B3,Haack}. Many new lines of research have also developed to consider further interesting generalisations. In \cite{B4}, Bhattacharyya \textit{et al} extended this result to nonrelativistic fluids and holographically obtained the incompressible nonrelativistic Navier-Stokes equations. Work has also been done on constructing the bulk duals of non-conformal fluids \cite{Skenderis3}, of charged fluids \cite{Erdmenger,Banerjee}, of superfluids \cite{Sonner,B5, Herzog}, and of anomalous fluids \cite{Majhi}.

A particularly interesting direction that will be the focus of this paper is the construction of bulk duals for forced fluid flows \cite{B2,Cai}. Solutions of fluid dynamics with particular forcing terms are known to exhibit
turbulence, which is a phenomenon that is not well understood. A holographic understanding of turbulence may well provide new insights on this topic. Research along these lines has already begun; some examples in the literature relating to holographic turbulence are \cite{Myers, Oz, Oz2, Liu3}. 

In this paper, we consider long wavelength solutions to the Einstein-dilaton system \textit{in arbitrary spacetime dimensions},
\begin{equation}
R_{AB} + d g_{AB} - \frac{1}{2}\partial_A \Phi \partial_B \Phi = 0,
\end{equation}
\begin{equation}
\nabla^2 \Phi = 0.
\end{equation}
These bulk metrics are dual to the forced fluid dynamical motions of boundary field theories
with actions of the form,
\begin{equation}  
S=\int\sqrt{g}e^{-\phi}\mathcal{L}.
\end{equation}
The boundary fluid obeys the following equations of motion,
\begin{equation} 
 \nabla_\mu T^{\mu\nu}=e^{-\phi} \mathcal{L} \nabla^\nu \phi,
\end{equation} 
which effectively are the relativistic fluid dynamical equations with an explicit dilaton-dependent forcing term. Stating our results more explicitly, we construct long wavelength, asymptotically locally AdS$_{d+1}$ bulk solutions with a slowly-varying boundary dilaton field and a weakly curved boundary metric to second order in a boundary derivative expansion. We also explicitly compute the fluid dynamical stress tensor and Lagrangian to second order in the derivative expansion thus generalising to arbitrary dimensions previous work by Bhattacharyya \textit{et al} \cite{B2} which was specific to a five-dimensional bulk spacetime.

This paper is organised as follows. In section \ref{2} we present a conceptual overview of the fluid/gravity correspondence highlighting the key steps in the construction of the bulk metric. Section \ref{3} then contains a review of the Weyl covariant notation for conformal fluid dynamics developed in \cite{Loganayagam} which we will be using throughout the rest of the paper. Our main results are contained in section \ref{4}; we present explicit solutions to the Einstein-dilaton equations valid for arbitrary spacetime dimensions to second order in a boundary derivative expansion, as well as expressions for the boundary stress tensor and Lagrangian accurate to the same order. Section \ref{5} has a discussion of our results and we end with two appendices giving more details of our calculations.

\section{Overview of the fluid/gravity correspondence}\label{2}
The aim of this section is to give a conceptual overview of the fluid/gravity correspondence. As we are primarily interested in conveying just the key ideas involved, we will focus on unforced fluid dynamics initially for simplicity. In the last subsection, we will explain how this methodology can be extended to holographic forced fluid dynamics by considering the Einstein-dilaton system. 

\subsection{A perturbative construction}

Up till this point, we have established that the Einstein equations with negative cosmological constant provide the dual dynamics of the stress tensor for an infinite class of strongly coupled field theories; and that if we are aiming to construct analytic, time-dependent holographic bulk solutions describing interesting, non-trivial, non-equilibrium phenomena, then fluid dynamics may well be a promising place to start. This intuition was spectacularly confirmed in \cite{B1}. Yet, given an arbitrary fluid dynamical configuration, how exactly would we go about explicitly constructing the dual bulk metric?

To probe this question further, it is useful to first examine the properties of fluid dynamics. These properties should be reflected in some analogous manner in the bulk solution. A deeper understanding of fluid dynamics therefore may well suggest an appropriate method of constructing the bulk dual. As presented thus far, fluid dynamics is a description of field theories at near-equilibrium subject to the constraint that the field theory must be locally equilibriated. But what does the condition of local
equilibrium imply? Suppose we have a field theory that is locally equilibriated: within patches of a certain size, say $\lambda_{equil}$ (which would be determined by the physics of the system\footnote{For example, for a dilute gas of weakly interacting particles, one would expect
$\lambda_{equil}$ to be of order the length of the mean free path.}), the system would have equilibriated, and it would be possible to assign meaningful values to thermodynamic quantities; a temperature, $T$, and a velocity, $u^\mu$, for instance. A temperature field, $T(x)$, and a velocity field, $u^\mu(x)$, can then be constructed by patching together these local values in some continuous sense. However, if a patch of size $\lambda_{equil}$ is required for a meaningful notion of equilibrium to be established (and for equilibrium variables to be assigned) then the temperature field, $T(x)$, and the velocity field, $u^\mu(x)$, necessarily can only vary on scales larger than $\lambda_{equil}$. They must be \textit{slowly-varying} with respect to the $\lambda_{equil}$ lengthscale\footnote{This applies in a temporal sense as well. The physics of the system concerned will also determine a characteristic timescale for equilibriation; the temporal variation of $T(x)$ and $u(x)$ must be slow relative to this.}. We call this the \textit{long wavelength limit}. Now, given that the fluid dynamical parameters must be slowly-varying, it is natural to therefore express the fluid dynamical stress tensor as an expansion in increasing order of derivatives, where higher order derivative terms are relatively suppressed compared to lower order terms. This fact should be emphasised. The condition of local equilibrium alone implies that fluid dynamics should naturally be expressed as an effective long wavelength theory, specified to some order in the derivative expansion. 

Now, given that the fluid dynamical stress tensor should be expressed in a derivative expansion, we must conclude that the bulk metric, which is dual to the boundary stress tensor, should correspondingly admit an expansion in \textit{boundary spacetime} derivatives,
\begin{equation} \label{MetricExpansion}
g_{AB} = g_{AB}^{(0)} + g_{AB}^{(1)} + g_{AB}^{(2)} + g_{AB}^{(3)} + \cdots,
\end{equation}
organised in increasing order of derivatives. In this sense, the bulk duals of fluid dynamics are approximate \textit{long wavelength} solutions to the Einstein equations with negative cosmological constant. With this in mind, we therefore see that to obtain the bulk metric dual to fluid dynamics, we should be aiming to solve the Einstein equations \textit{perturbatively} to some specified accuracy in the boundary derivative expansion.

\subsection{Slow variation of bulk tubes and the zeroth order ansatz}

For the simplest case of holographic unforced fluid dynamics, we have to solve the following equations perturbatively,
\begin{equation} \label{Einstein}
E_{AB} \equiv R_{AB} - \frac{1}{2}Rg_{AB} + \Lambda g_{AB} = 0, \;\;\;\;\Lambda \equiv -\frac{d(d-1)}{2}.
\end{equation}
However, we are now confronted with the question: what should we choose as our zeroth order ansatz, $g_{AB}^{(0)}$?

We know that the AdS-Schwarzschild metric\footnote{Here, $r$ represents the bulk radial coordinate while $x^\mu$ labels the boundary coordinates.},
\begin{equation}
\begin{aligned}
ds^2&=\frac{dr^2}{r^2 f(br)} + r^2 (-f(br) u_\mu u_\nu dx^\mu dx^\nu + P_{\mu\nu}dx^\mu dx^\nu),  \\
f(br)&=1-\frac{1}{(br)^d}, \quad \eta_{\mu\nu}u^\mu u^\nu = -1, \quad P_{\mu\nu}=\eta_{\mu\nu}+u_\mu u_\nu, \quad b= \frac{d}{4\pi T},
\end{aligned} 
\end{equation}
is dual to a field theory state in global equilibrium with temperature $T$ and velocity $u^{\mu}$. Given that fluid dynamics describes field theory configurations which are locally equilibriated, it is not unreasonable to propose that the dual bulk solution should approximately be given by patching together tubes of AdS black brane solutions with different values for $T$ and $u^{\mu}$. A natural first guess for the zeroth order ansatz could therefore be:
\begin{equation}\label{Schwarzschild}
\begin{aligned}
ds^2&=\frac{dr^2}{r^2 f(b(x)r)} + r^2 (-f(b(x)r) u_\mu (x)u_\nu (x) dx^\mu dx^\nu + P_{\mu\nu}(x)dx^\mu dx^\nu),  \\
P_{\mu\nu}(x)&=g_{\mu\nu}(x)+u_\mu (x) u_\nu (x) , \quad b (x) = \frac{d}{4\pi T(x)},
\end{aligned} 
\end{equation}
where $g_{\mu\nu}(x)$, $b(x)$, and $u^{\mu}(x)$ are all slowly-varying functions of the boundary coordinates. Locally (in the field theory directions), this metric is indistinguishable from a uniform black brane metric, but the values for $T$ and $u^{\mu}$ change as we move along the boundary.

It turns out that this guess is actually incorrect, and the reason for this is quite subtle. The problem lies in how the tubes of uniform black brane solutions (which follow lines of constant $x^\mu$) extend from the boundary into the bulk. We will now elaborate on this issue.

\begin{figure}
 \begin{center}
\includegraphics[scale=0.8]{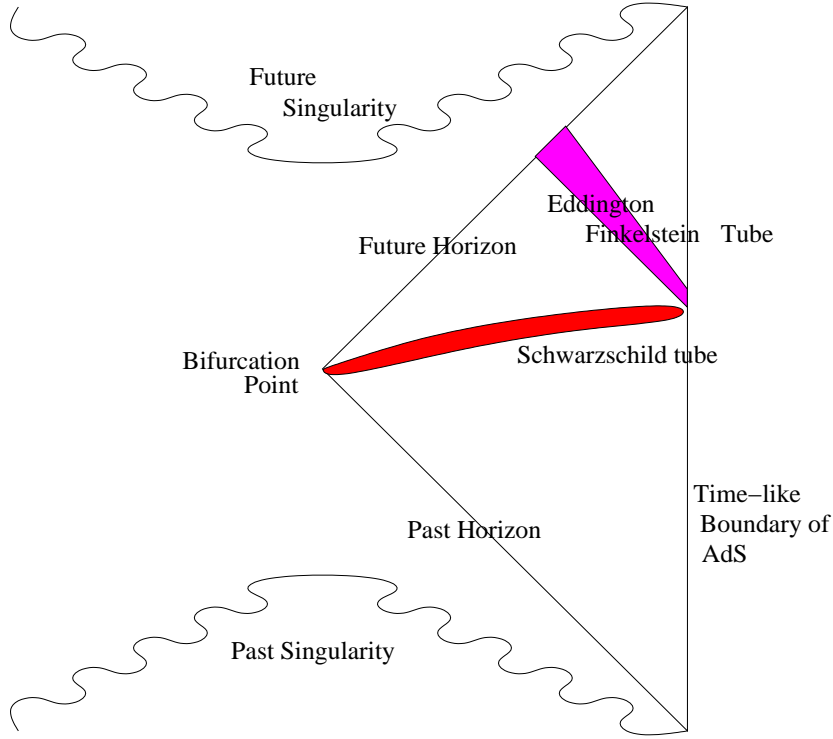}
\end{center}
\caption{This shows the Penrose diagram of a uniform black brane; tubes of constant $x^\mu$ are shown in both the Eddington-Finkelstein and Schwarzschild coordinates. This figure is taken from \cite{B3}.}
\label{figure}
\end{figure}

For any arbitrary fluid dynamical configuration, if a sudden forcing were applied at some coordinate $y^\mu$, then only a certain region (within the future boundary light cone of $y^\mu$) will be affected by this forcing. We refer to this region as $C(y^\mu)$. Now, let us consider the bulk region (which we refer to as $B(y^\mu)$) that consists of the union of all tubes which stem from $C(y^\mu)$. The region $B(y^\mu)$ will certainly feel the effects of the forcing at $y^\mu$. And therefore, by causality, $B(y^\mu)$ must be completely contained within the future bulk light cone of $y^\mu$. This condition is not satisfied if we use the metric (\ref{Schwarzschild}) which is written in Schwarzschild coordinates (see figure \ref{figure}). If we instead rewrite our black brane solutions in Eddington-Finkelstein coordinates, 
\begin{equation}\label{EF}
ds^2=-2u_\mu (x) dx^\mu dr - r^2f(b(x) r) u_\mu (x) u_\nu (x) dx^\mu dx^\nu + r^2 P_{\mu\nu} (x) dx^\mu dx^\nu,
\end{equation}
allowing the temperature $T(x)$ (or equivalently $b(x)$), the velocity $u(x)$, and the boundary metric $g_{\mu\nu} (x)$ to be slowly-varying boundary functions, we find that these tubes will run along ingoing null geodesics and that there will no longer be any issue with causality. It therefore seems appropriate to input the metric (\ref{EF}) as our zeroth order ansatz as we perturbatively solve the Einstein equations.

\subsection{Equations at each order in the boundary derivative expansion}

In this subsection, we examine the structure of the equations that result as we attempt to solve the Einstein equations perturbatively. The remarkable simplification which occurs in the long wavelength limit is what enables us to obtain analytic solutions to the Einstein equations without sacrificing nonlinearity.

Rewriting the metric as an expansion in boundary derivatives (\ref{MetricExpansion}) and plugging this into the gravitational equations (\ref{Einstein}), we arrive at the following schematic form of equations\footnote{We have suppressed the spacetime indices in the metric, $g$, for notational convenience.}:
\begin{equation} \label{Schematic}
\mathbb{H}\left[g^{(0)}\right] g^{(n)}(r,x^\mu) = s_n.
\end{equation}
We have focused only on terms of order $n$ in boundary derivatives. The assumption is that all $g^{(m)}$ for $m\le n-1$ have already been determined by the perturbation theory at lower orders; $g^{(n)}$ is the only unknown function at this order. We now highlight several important properties of the differential operator $\mathbb{H}$ (which acts on $g^{(n)}$) and the source terms $s_n$.

Observe that because $g^{(n)}$ is already of order $n$ in boundary derivatives, the differential operator is necessarily linear. Further, it must be a differential operator only in $r$ and all coefficients must be zeroth order functions. In this sense, $\mathbb{H}$ is ultralocal in the field theory directions; it cannot have any boundary derivatives. We also note that $\mathbb{H}$ is a second order differential operator which is the same at each order in the perturbation theory. It is second order because it inherits the structure of the Einstein equations, and it is independent of $n$ because the exact same combinations of zeroth order functions and partial derivatives in $r$ which act on $g^{(n)}$ will also act on $g^{(m)}$ for all $m\ge 1$, and hence we must have the same homogeneous operator at all orders in the perturbation theory. The source terms, however, which consist of boundary derivatives acting on lower order functions, will be different at each order.

It is now certainly worth pausing to emphasise what has been achieved. In this long wavelength limit, the Einstein equations have reduced to a system of inhomogeneous second order linear differential equations in the variable $r$ alone. Here we again stress that we have not sacrificed the nonlinearity of the Einstein equations; the fact that the differential operator is linear is an advantage that we obtain by working perturbatively order by order in boundary derivatives. This deceptive linearity, coupled with the ultralocality in the boundary directions, is what makes the Einstein equations so much more tractable in this long wavelength limit.

We now comment further on the nature of these equations. As we are working in $d+1$ spacetime dimensions, the system (\ref{Schematic}) will provide us with $\frac{(d+1)(d+2)}{2}$ equations. Only $\frac{d^2 + d + 2}{2}$ of these equations will explicitly involve the unknown function $g^{(n)}$; and of these, one will prove to be redundant. We refer to these equations as the `dynamical' equations. The remaining $d$ will only involve boundary derivatives of lower order terms, $g^{(m)}$ for $m\le n-1$. We refer to these as the `constraint' equations. 

The dynamical equations can always be solved by direct integration for an arbitrary source $s_n$; and subject to imposing regularity at $r>0$ and normalisability at infinity, a unique\footnote{There is a further ambiguity associated with redefinitions of the velocity and temperature fields, but this can be fixed by a choice of convention. This is explained in detail in \cite{B1}.} solution can be obtained. The constraint equations, on the other hand, impose relations between boundary derivatives of $g^{(m)}$ for $m\le n-1$. And since these $g^{(m)}$ are themselves constructed from appropriate derivatives of the velocity ($u^\mu(x)$) and temperature ($T(x)$) fields, the constraint equations are ultimately relations constraining the allowed forms of $u^\mu(x)$ and $T(x)$. These constraint equations have an especially simple boundary interpretation; they are the equations of conservation of the boundary stress tensor at one order lower,
\begin{equation}
\nabla_\mu T^{\mu\nu}_{(n-1)} = 0.
\end{equation}
And in this particular long wavelength limit, these equations are found to be equivalent to the equations of conformal relativistic fluid dynamics for the distinguished fluid dual to Einstein gravity. This concretely shows that the perturbative procedure outlined above correctly produces the class of bulk solutions dual to fluid dynamics. It should thus be emphasised that the fluid/gravity correspondence constitutes an explicit proof of the gauge/gravity duality in this long wavelength limit.

\subsection{Extension to the Einstein-dilaton system}

In this final subsection, we explain how the methodology previously described can be extended to the Einstein-dilaton system to produce bulk solutions dual to forced fluid dynamics.

The Einstein-dilaton system is governed by the following equations:
\begin{equation} 
\begin{aligned}
E_{AB}^\Phi & \equiv R_{AB} - \frac{1}{2}Rg_{AB} + \Lambda g_{AB} - \frac{1}{2}\partial_A \Phi \partial_B \Phi + \frac{1}{4} \left(\partial \Phi \right)^2 g_{A B}= 0, \\
\nabla^2 \Phi & = 0.
\end{aligned}
\end{equation}
And upon resubstituting for the Ricci scalar $R$ and the cosmological constant $\Lambda$, this then simplifies to:
\begin{equation}\label{Einstein-dilaton}
\begin{aligned}
E_{AB}^\Phi &\equiv R_{AB} + d g_{AB} - \frac{1}{2}\partial_A \Phi \partial_B \Phi = 0,\\
\nabla^2 \Phi & = 0.
\end{aligned}
\end{equation}
By using standard holographic formulae for the stress tensor $T_{\mu\nu}$ and the Lagrangian $\mathcal{L}$ and by analysing appropriate projections of the bulk equations near the boundary (see Appendix A of \cite{B2}), it can be shown directly that the boundary dynamics dual to the Einstein-dilaton system are given by:
\begin{equation} \label{ForcedFluid}
\nabla_\mu T^{\mu\nu} = e^{-\phi} \mathcal{L} \nabla^\nu \phi,
\end{equation}
where $\phi$ is the projection of $\Phi$ onto the boundary. In the long wavelength limit, we can therefore view the Einstein-dilaton system as being the dual bulk dynamics for conformal relativistic fluid dynamics with a dilaton-dependent forcing term.

We now proceed in an analogous manner to what was done previously. Field theory intuition again tells us that the metric should be a slowly-varying function in the boundary coordinates and that we should thus be aiming to solve the equations (\ref{Einstein-dilaton}) perturbatively to a certain accuracy in boundary derivatives. However, for the Einstein-dilaton system, the metric couples to the dilaton, and so we must further require that the dilaton be slowly-varying in the boundary directions as well. Hence, it must also admit an expansion in boundary derivatives:
\begin{equation}
\Phi=\Phi^{(0)} + \Phi^{(1)} + \Phi^{(2)} + \Phi^{(3)} + \cdots.
\end{equation}

We must again address the issue of choosing the zeroth order ansatz, but this is just a straightforward generalisation of the unforced case. The Einstein-dilaton system admits uniform black brane solutions of the following form,
\begin{equation}
\begin{aligned}
ds^2&=-2u_\mu  dx^\mu dr - r^2f(b r) u_\mu  u_\nu  dx^\mu dx^\nu + r^2 P_{\mu\nu}  dx^\mu dx^\nu, \\
\Phi &= \phi_0, 
\end{aligned}
\end{equation}
where $\phi_0$ is a constant. Patching together tubes of uniform black brane solutions with different parameter values gives us our zeroth order ansatz\footnote{We will actually be using a Weyl-covariant form of this ansatz; we will elaborate on this in section \ref{3}.}:
\begin{equation}
\begin{aligned}
ds^2&=-2u_\mu (x) dx^\mu dr - r^2f(b(x) r) u_\mu (x) u_\nu (x) dx^\mu dx^\nu + r^2 P_{\mu\nu} (x) dx^\mu dx^\nu, \\
\Phi &= \phi(x).
\end{aligned}
\end{equation}

With all of this in hand, we can now substitute the expansions for the metric and the dilaton into the equations (\ref{Einstein-dilaton}) and examine the structure of the resulting equations. The equations at order $n$ in the derivative expansion can be schematically represented as:
\begin{equation} \label{ForcedMetric}
\mathbb{H}\left[g^{(0)}\right] g^{(n)}(r,x^\mu) = s_n,
\end{equation}
\begin{equation} \label{ForcedDilaton}
\mathbb{H}^\Phi\left[g^{(0)}\right] \Phi^{(n)}(r,x^\mu)=s^\Phi_n,
\end{equation}
where $\mathbb{H}^\Phi$ and $s^\Phi_n$ are the differential operator and source terms for $\Phi$ respectively. The dynamical equations of (\ref{ForcedMetric}) together with the equation for the dilaton (\ref{ForcedDilaton}) are sufficient to determine $g^{(n)}$ and $\Phi^{(n)}$. The remaining $d$ constraint equations reduce to the equations of forced fluid dynamics (\ref{ForcedFluid}).

\section{Manifest Weyl covariance}\label{3}

In this section, we review the Weyl covariant formalism introduced in \cite{Loganayagam} for conformal relativistic fluid dynamics. This formalism allows for more compact notation. Also, as we shall see in the final subsection, the components of the bulk metric can be classified according to how they transform under Weyl rescaling; thus, it is convenient to adopt a formalism which makes their Weyl transformation properties manifest.

\subsection{Regulation and Weyl symmetry}

The aim of this subsection is to elaborate on a well-known subtlety in the AdS/CFT correspondence relating to the interpretation of the boundary field theory. This subtlety in interpretation leads to the Weyl covariant nature of the boundary fluid dynamics.

We begin by noting that to obtain the dual field theory interpretation of a bulk solution, one needs to regulate the solution near the boundary on slices of constant $r$, the radial coordinate. More concretely, the bulk solution will be interpreted as a state of the dual field theory on a background whose metric is related to the induced metric on the regulated boundary. However, there is a well-known ambiguity in the choice of the radial coordinate. To illustrate this further, consider the following parametrisation of AdS:
\begin{equation} \label{AdS}
ds^2=-2 u_{\mu} dx^\mu dr + r^2 g_{\mu\nu} dx^\mu dx^\nu.
\end{equation}
With this choice of coordinates, the dual field theory is considered to live on a background whose metric is given precisely by $g_{\mu\nu}$. If we instead choose a different radial coordinate $\tilde{r}$, given by a constant rescaling of $r$, and replace $g_{\mu\nu}$ and $u_\mu$ as follows:
\begin{equation} 
r=\lambda^{-1}\tilde{r}, \quad u_\mu=\lambda \tilde{u}_\mu, \quad g_{\mu\nu}=\lambda^2 \tilde{g}_{\mu\nu},
\end{equation}
for constant $\lambda$, the bulk metric takes the following (invariant) form:
\begin{equation}
ds^2=-2 \tilde{u}_{\mu} dx^\mu d\tilde{r} + \tilde{r}^2 \tilde{g}_{\mu\nu} dx^\mu dx^\nu.
\end{equation}
Regulating on surfaces of constant $\tilde{r}$ gives us a field theory vacuum state on a background $\tilde{g}_{\mu\nu}=\lambda^{-2} g_{\mu\nu}$. This equivalence between boundary metrics related by a constant rescaling arises from the dilatational symmetry of AdS, $SO(1,1)$. The full symmetry group of AdS, however, is the conformal group, $SO(d,2)$, and although this symmetry isn't explicitly manifest in the choice of coordinates (\ref{AdS}), the bulk AdS spacetime must therefore be dual to a field theory state defined on a space with any of the infinite number of metrics Weyl equivalent to $g_{\mu\nu}$; this reflects the Weyl symmetry of the dual field theory.

Now, bulk spacetimes dual to fluid dynamics are asymptotically locally AdS. As such, the boundary fluid dynamics should correspondingly be Weyl invariant. However, in contrast to AdS spacetime (\ref{AdS}), this boundary Weyl symmetry is explicitly manifest in the bulk metric. By this we mean that if we choose to regulate the fluid dynamical bulk spacetime using a locally rescaled radial coordinate, $r=e^{-\chi(x^\mu)} \tilde{r}$, and perform the following simultaneous replacements:
\begin{equation} \label{Replacements}
r=e^{-\chi} \tilde{r}, \quad u_\mu=e^{\chi} \tilde{u}_\mu, \quad b=e^{\chi} \tilde{b}, \quad g_{\mu\nu}=e^{2\chi} \tilde{g}_{\mu\nu},
\end{equation}
the form of the bulk metric will remain invariant. We will now proceed to prove this. As previously established, bulk spacetimes dual to fluid dynamics admit an expansion in boundary derivatives of the form,
\begin{equation}  \label{Initial}
g_{AB} = g_{AB}^{(0)} + g_{AB}^{(1)} + g_{AB}^{(2)} + g_{AB}^{(3)} + \cdots,
\end{equation}
where the zeroth order contribution, $g_{AB}^{(0)}$, is given by:
\begin{equation}
ds^2=-2u_\mu dx^\mu dr - r^2f(b r) u_\mu  u_\nu  dx^\mu dx^\nu + r^2 P_{\mu\nu}  dx^\mu dx^\nu.
\end{equation}
Now, if we perform the simultaneous replacements (\ref{Replacements}), the bulk metric will take the form:
\begin{equation}
\tilde{g}_{AB} = \tilde{g}_{AB}^{(0)} + \tilde{g}_{AB}^{(1)} + \tilde{g}_{AB}^{(2)} + \tilde{g}_{AB}^{(3)} + \cdots,
\end{equation}
where the terms are functions of the new rescaled variables:
\begin{equation}
 \tilde{g}_{AB}^{(n)}\equiv\tilde{g}_{AB}^{(n)}(\tilde{r}, \tilde{u}_\mu, \tilde{b}) \quad \forall n.
\end{equation}
But note that the form of the zeroth order contribution, $\tilde{g}_{AB}^{(0)}$, remains invariant under (\ref{Replacements}), \textit{i.e.}:
\begin{equation} \label{Initialdata}
\begin{aligned}
ds^2&=-2\tilde{u}_\mu dx^\mu d\tilde{r} - \tilde{r}^2f(\tilde{b} \tilde{r}) \tilde{u}_\mu  \tilde{u}_\nu  dx^\mu dx^\nu + \tilde{r}^2 \tilde{P}_{\mu\nu}  dx^\mu dx^\nu, \\
\tilde{P}_{\mu\nu}&=\tilde{g}_{\mu\nu}+\tilde{u}_\mu \tilde{u}_\nu.
\end{aligned}
\end{equation}
However, we equally could have directly used the expression (\ref{Initialdata}) as our zeroth order ansatz to perturbatively construct a bulk spacetime with (\ref{Initialdata}) as the fluid dynamical initial data. In doing so, we would have obtained a bulk spacetime identical to (\ref{Initial}) at each order except with the variables $r, u_\mu,$ and $b$ replaced by $\tilde{r}, \tilde{u}_\mu,$ and $\tilde{b}$. Now, recall that our perturbative procedure constructs a \textit{unique} bulk spacetime for a specified zeroth order ansatz. Hence we are forced to conclude that the bulk spacetime obtained from (\ref{Initial}) by performing the simultaneous replacements (\ref{Replacements}) must be the same as the bulk spacetime obtained directly from the zeroth order ansatz (\ref{Initialdata}); this would be identical to (\ref{Initial}) at all orders except with $r, u_\mu,$ and $b$ replaced by $\tilde{r}, \tilde{u}_\mu,$ and $\tilde{b}$. This concludes our proof; the bulk spacetime dual to fluid dynamics is therefore invariant under the simultaneous replacements (\ref{Replacements}). 

Observe that, from the perspective of the boundary, (\ref{Replacements}) is nothing more than a boundary Weyl transformation with $u_\mu$ and $b$ transforming as Weyl tensors of weight $-1$. It is thus convenient to adopt a Weyl covariant formalism; we develop this further in the next subsection.

\subsection{Weyl covariant derivative}
A Weyl covariant tensor is a quantity that transforms homogeneously under a Weyl transformation. More specifically, a tensor of weight $w$ transforms as follows:
\begin{equation}
\mathcal{Q}^{\mu\cdots}_{\nu\cdots}=e^{-w\chi(x)}\tilde{\mathcal{Q}}^{\mu\cdots}_{\nu\cdots}
\end{equation}
under a Weyl rescaling, $g_{\mu\nu}=e^{2\chi(x)}\tilde{g}_{\mu\nu}$.
The main obstruction to maintaining explicit Weyl covariance is that ordinary covariant derivatives of Weyl covariant tensors are not themselves Weyl covariant. This problem can be circumvented by introducing a `Weyl covariant derivative'; this was the main technical innovation of \cite{Loganayagam}. The action of the Weyl covariant derivative on an arbitrary tensor $\mathcal{Q}^{\mu\cdots}_{\nu\cdots}$ of weight $w$ is defined by:
\begin{equation} \label{Weylderivative}
\begin{aligned}
\mathcal{D}_\lambda \mathcal{Q}^{\mu\cdots}_{\nu\cdots} &\equiv \nabla_\lambda \mathcal{Q}^{\mu\cdots}_{\nu\cdots} + w \mathcal{A}_\lambda \mathcal{Q}^{\mu\cdots}_{\nu\cdots} \\
&\quad + \left[g_{\lambda\alpha}\mathcal{A}^\mu-\delta_\lambda^\mu\mathcal{A}_\alpha - \delta_\alpha^\mu \mathcal{A}_\lambda \right] \mathcal{Q}^{\alpha\cdots}_{\nu\cdots} +\cdots \\
&\quad - \left[g_{\lambda\nu}\mathcal{A}^\alpha-\delta_\lambda^\alpha\mathcal{A}_\nu - \delta_\nu^\alpha \mathcal{A}_\lambda \right] \mathcal{Q}^{\mu\cdots}_{\alpha\cdots} -\cdots.
\end{aligned}
\end{equation}
The Weyl connection, $\mathcal{A}_\mu$, is constructed from the fluid velocity field, $u_\mu$, as follows:
\begin{equation}
\mathcal{A}_\mu \equiv u^\lambda \nabla_\lambda u_\mu - \frac{\nabla_\lambda u^\lambda}{d-1} u_\mu = \tilde{\mathcal{A}}_\mu + \partial_\mu \chi.
\end{equation}
As we can see from the last equality, this expression for $\mathcal{A}_\mu$ transforms in a similar manner to a metric connection under a Weyl transformation. This is what enables us to construct a derivative that is Weyl covariant, as done in (\ref{Weylderivative}); the parts of  $\nabla_\lambda \mathcal{Q}^{\mu\cdots}_{\nu\cdots}$ which do not transform homogeneously are cancelled by the terms involving $\mathcal{A}_\mu$. It can further be shown that the Weyl covariant derivative of a tensor of weight $w$ is itself a tensor of weight $w$.

We will now introduce several Weyl covariant tensors that will be used throughout the rest of the paper. The following tensors are naturally constructed from the Weyl covariant derivative:
\begin{equation}
\begin{aligned}
\left[\mathcal{D}_\mu, \mathcal{D}_\nu\right]V_\lambda &\equiv w\mathcal{F}_{\mu\nu} V_\lambda + {\mathcal{R}_{\mu\nu\lambda}}^\alpha V_\alpha \quad \text{with} \\
\mathcal{F}_{\mu\nu} &\equiv \nabla_\mu \mathcal{A}_\nu - \nabla_\nu \mathcal{A}_\mu \quad \text{and}\\
\mathcal{R}_{\mu\nu\lambda\sigma} &\equiv R_{\mu\nu\lambda\sigma} + 
\mathcal{F}_{\mu\nu} g_{\lambda\sigma} - \delta^\alpha_{[\mu } g_{\nu ] [ \lambda} \delta^\beta_{\sigma ]} \left(\nabla_\alpha\mathcal{A}_\beta + \mathcal{A}_\alpha \mathcal{A}_\beta - \frac{\mathcal{A}^2}{2} g_{\alpha\beta} \right).
\end{aligned}
\end{equation}
We will also make use of the following two contractions obtained from the Weyl covariant Riemann tensor, $\mathcal{R}_{\mu\nu\lambda\sigma} $:
\begin{equation}
\begin{aligned}
\mathcal{R}_{\mu\nu} &\equiv {\mathcal{R}_{\mu\lambda\nu}}^\lambda = R_{\mu\nu} + (d-2)(\nabla_\mu\mathcal{A}_\nu + \mathcal{A}_\mu\mathcal{A}_\nu-\mathcal{A}^2 g_{\mu\nu} ) + g_{\mu\nu} \nabla_\lambda \mathcal{A}^\lambda + \mathcal{F}_{\mu\nu}; \\
\mathcal{R} &\equiv {\mathcal{R}_\lambda}^\lambda = R + 2(d-1) \nabla_\lambda \mathcal{A}^\lambda - (d-2)(d-1)\mathcal{A}^2.
\end{aligned}
\end{equation}
And finally we define the shear strain rate, $\sigma_{\mu\nu}$, and vorticity, $\omega_{\mu\nu}$, of the boundary fluid:
\begin{equation}
\begin{aligned}
\sigma_{\mu\nu} &\equiv \mathcal{D}_{( \mu} u_{\nu ) }; \\
\omega_{\mu\nu} &\equiv \mathcal{D}_{[\mu} u_{\nu ] } .
\end{aligned}
\end{equation}

\subsection{Independent Weyl invariant tensors}
Here, we classify all Weyl invariant scalars, transverse\footnote{By transverse we mean orthogonal to $u^\mu$.} vectors, and symmetric traceless transverse tensors up till second order in derivatives; this will be of importance in the following subsection. There are two subtleties involved in this classification that we should first highlight. Note that the equations of motion, $\nabla_\mu T^{\mu\nu} = e^{-\phi} \mathcal{L} \nabla^\nu \phi$, impose relations between various Weyl covariant quantities; thus, in our counting, we only list Weyl tensors which are independent on-shell. And also, since the dilaton is Weyl invariant, any Weyl invariant tensor can be multiplied by a function of $\phi$ to get another independent Weyl invariant quantity; we will neglect this complication in our classification as well. 

We begin with the zeroth order Weyl invariant tensors. We aim to construct Weyl invariants using the boundary dilaton field, $\phi$, the boundary metric, $g_{\mu\nu}$, and the fluid dynamical quantities, $b$ and $u_\mu$. The boundary dilaton $\phi$ is a Weyl invariant scalar while $b$ and $u_\mu$ transform homogeneously under Weyl rescalings with weight $-1$. It thus follows that there are no nontrivial Weyl invariant scalars, transverse vectors, or symmetric traceless transverse tensors at zeroth order in derivatives.

To obtain the Weyl invariants at first order, we must consider the first order relations imposed by the equations of motion, $\nabla_\mu T^{\mu\nu} = e^{-\phi} \mathcal{L} \nabla^\nu \phi$. It is easy to see that these relations arise from the zeroth order contributions to the stress tensor, $T^{\mu\nu}$, and Lagrangian, $\mathcal{L}$. For the stress tensor, the zeroth order contribution is simply that of a perfect fluid, $b^{-d}\left(g^{\mu\nu} + d u^\mu u^\nu \right)$. And for the Lagrangian, there can be no zeroth order contribution. The reason for this is as follows: if we set $\phi$ to be a constant, the Einstein-dilaton system must consistently truncate to the Einstein equations with negative cosmological constant. Correspondingly, the boundary fluid dynamics must reduce to that of the unforced case. Thus, the lowest order contribution to the Lagrangian must be proportional to a derivative of $\phi$; there can be no zeroth order terms. Analysing the resulting first order relations, it can be shown that first order partial derivatives of $b$ can be expressed as derivatives of $u_\mu$\footnote{Please see Appendix C of \cite{B3} for a more detailed explanation of this.}. It follows that there is only one independent Weyl invariant scalar at first order $\left(\text{which can be taken to be }b u^\mu \mathcal{D}_\mu \phi \right)$, one Weyl invariant transverse vector $\left(P^\nu_\mu \mathcal{D}_\nu \phi\right)$, and one Weyl invariant symmetric traceless transverse tensor $\left(b^{-1} \sigma_{\mu\nu} \right)$.

For the second order Weyl invariant tensors, we must similarly consider the relations imposed at second order by the equations of motion; these originate from the first order contributions to $T^{\mu\nu}$ and $\mathcal{L}$. The stress tensor, $T^{\mu\nu}$, transforms with weight $d+2$; and thus, using our previous classification of first order Weyl invariants, we can deduce that the first order contribution to $T^{\mu\nu}$ must be proportional to $b^{1-d} \sigma^{\mu\nu}$. And for the Lagrangian, which transforms with weight $d$, we can similarly conclude that the first order term must be of the form $b^{1-d} u^\mu \mathcal{D}_\mu \phi $. The two derivative relations which result can be used to express the partial derivatives of $b$ to second order in terms of derivatives of $u_\mu$ and $\phi$. With all of this in hand, it is not too difficult to show that there are seven independent Weyl invariant scalars:
\begin{equation}
\begin{aligned}
&b^2\sigma_{\mu\nu}\sigma^{\mu\nu}, \quad b^2\omega_{\mu\nu}\omega^{\mu\nu}, \quad b^2 \mathcal{R}, \\
b^2 P^{\mu\nu}\mathcal{D}_\mu \mathcal{D}_\nu \phi, \quad b^2 u^\mu u^\nu &\mathcal{D}_\mu \mathcal{D}_\nu \phi, \quad b^2 P^{\mu\nu} \mathcal{D}_\mu \phi \mathcal{D}_\nu \phi, \quad \text{and} \quad b^2 u^\mu u^\nu \mathcal{D}_\mu \phi \mathcal{D}_\nu \phi,
\end{aligned}
\end{equation}
six Weyl invariant transverse vectors:
\begin{equation}
\begin{aligned}
b P^\nu_\mu \mathcal{D}_\lambda {\sigma_\nu}^\lambda, \quad &b P^\nu_\mu \mathcal{D}_\lambda {\omega_\nu}^\lambda, \quad b P^\nu_\mu u^\lambda \mathcal{D}_\nu \mathcal{D}_\lambda \phi, \quad b P^\nu_\mu u^\lambda \mathcal{D}_\nu \phi \mathcal{D}_\lambda \phi, \\
&b {\sigma_\mu}^\lambda \mathcal{D}_\lambda \phi, \quad \text{and} \quad b {\omega_\mu}^\lambda \mathcal{D}_\lambda \phi,
\end{aligned}
\end{equation}
and eight Weyl invariant symmetric traceless transverse tensors:
\begin{equation}
\begin{aligned}
u^\lambda \mathcal{D}_\lambda \sigma_{\mu\nu}, \quad \sigma_{\mu\nu} u^\lambda \mathcal{D}_\lambda \phi, \quad &C_{\mu\alpha\nu\beta} u^\alpha u^\beta, \quad {\omega_\mu}^\lambda \sigma_{\lambda\nu} + {\omega_\nu}^\lambda \sigma_{\lambda\mu}, \\
\frac{1}{2} \left[ P^\alpha_\mu P^\beta_\nu + P^\alpha_\nu P^\beta_\mu - \frac{2}{d-1} P^{\alpha\beta} P_{\mu\nu} \right] &\mathcal{D}_\alpha \mathcal{D}_\beta \phi, \quad  
\left[ P^\alpha_\mu P^\beta_\nu - \frac{1}{d-1} P^{\alpha\beta} P_{\mu\nu} \right] \mathcal{D}_\alpha \phi \mathcal{D}_\beta \phi, \\
{\sigma_\mu}^\lambda \sigma_{\lambda\nu} - \frac{1}{d-1} P_{\mu\nu} \sigma_{\alpha\beta} \sigma^{\alpha\beta}&, \quad \text{and} \quad {\omega_\mu}^\lambda \omega_{\lambda\nu} + \frac{1}{d-1} P_{\mu\nu} \omega_{\alpha\beta} \omega^{\alpha\beta}.
\end{aligned}
\end{equation}

\subsection{Weyl covariant form of the fluid dynamical metric}
In this final subsection, we demonstrate that it is possible to use boundary Weyl invariance to constrain the form of the bulk metric. In more detail, we show that because the bulk metric is invariant under the simultaneous replacements (\ref{Replacements}), the components of the bulk metric can be classified according to how they transform under boundary Weyl rescalings.

Before we proceed further, we must first choose a gauge for the bulk metric. We use the same gauge\footnote{Some early work on the fluid/gravity correspondence \cite{B1,B2} used a different gauge, given by:
\begin{equation}
g_{rr}=0, \quad g_{r\mu} \propto u_\mu, \quad Tr\left(\left(g^{(0)}\right)^{-1}g^{(m)}\right)=0\quad(m>0).
\end{equation}
All of our results can be recast in this gauge by making an appropriate change of variables.}
 as \cite{B3}, which is specified by:
\begin{equation} \label{gauge}
g_{rr}=0, \quad g_{r\mu}=-u_\mu.
\end{equation}
This gauge has the nice geometric interpretation that lines of constant $x^\mu$ are ingoing null geodesics with $r$ being an affine parameter along them. Also, note that this gauge choice is invariant under the transformation (\ref{Replacements}). 

Now, observe that, consistent with our gauge choice (\ref{gauge}), we can parametrise our bulk metric as follows:
\begin{equation} \label{InitialBulk}
ds^2=-2u_\mu dx^\mu \left( dr + \mathcal{V}_\nu\left(r,u_\alpha, b\right) dx^\nu\right) + \mathcal{G}_{\mu\nu} \left(r,u_\alpha, b\right) dx^\mu dx^\nu \quad \text{with } \mathcal{G}_{\mu\nu} \text{ transverse.}
\end{equation}
We aim to determine how the functions $\mathcal{V}_\nu$ and $\mathcal{G}_{\mu\nu}$ transform under (\ref{Replacements}) which effectively is a boundary Weyl transformation. Recall that the fluid dynamical bulk metric is invariant under the simultaneous replacements (\ref{Replacements}), thus, under this transformation, the bulk metric (\ref{InitialBulk}) becomes:
\begin{equation} \label{FinalBulk}
\begin{aligned}
ds^2&=-2\tilde{u}_\mu dx^\mu \left( d\tilde{r} + \mathcal{V}_\nu\left(\tilde{r},\tilde{u}_\alpha, \tilde{b}\right) dx^\nu\right) + \mathcal{G}_{\mu\nu} \left(\tilde{r},\tilde{u}_\alpha, \tilde{b}\right) dx^\mu dx^\nu \\
&= -2u_\mu dx^\mu \left( dr + e^{-\chi} \mathcal{V}_\nu\left(\tilde{r},\tilde{u}_\alpha, \tilde{b}\right) dx^\nu + r \partial_\nu \chi dx^\nu \right) + \mathcal{G}_{\mu\nu} \left(\tilde{r},\tilde{u}_\alpha, \tilde{b}\right) dx^\mu dx^\nu.
\end{aligned}
\end{equation}
By comparing the two equivalent metrics (\ref{InitialBulk}) and (\ref{FinalBulk}), we can deduce the transformation properties of $\mathcal{V}_\nu$ and $\mathcal{G}_{\mu\nu}$:
\begin{equation}
\mathcal{V}_\nu\left(r,u_\alpha, b\right)=e^{-\chi} \left[ \mathcal{V}_\nu\left(\tilde{r},\tilde{u}_\alpha, \tilde{b}\right) + \tilde{r} \partial_\nu \chi \right] \quad \text{and} \quad \mathcal{G}_{\mu\nu} \left(r,u_\alpha, b\right)=\mathcal{G}_{\mu\nu} \left(\tilde{r},\tilde{u}_\alpha, \tilde{b}\right).
\end{equation}
It follows that $\mathcal{V}_\nu - r \mathcal{A}_\nu$ must be a linear sum of Weyl covariant vectors (both transverse and non-transverse) of weight $+1$ with coefficients that are arbitrary functions of $br$. Similarly, $\mathcal{G}_{\mu\nu}$ must be a linear sum of Weyl invariant tensors. These Weyl covariant vectors of weight $+1$ and the Weyl invariant tensors can easily be obtained from our classification in the previous subsection. The functions of $br$, however, must be determined by direct calculation.

In keeping with explicit Weyl covariance, we now choose a slightly different starting ansatz, $g_{AB}^{(0)}$:
\begin{equation}
ds^2=-2u_\mu dx^\mu \left( dr + \left(r\mathcal{A}_\nu+ \frac{r^2 f(br)}{2} u_\nu\right) dx^\nu\right) + r^2 P_{\mu\nu}  dx^\mu dx^\nu.
\end{equation}
Using this ansatz, we can perturbatively solve the Einstein equations and determine the functions $\mathcal{V}_\nu$ and $\mathcal{G}_{\mu\nu}$ to any order in boundary derivatives. We present the results of such a calculation to second order in the next section.

\section{Explicit results up to second order}\label{4}
Here, we present our results for the fluid dynamical bulk metric $g_{AB}$ and the dilaton $\Phi$, as well as the corresponding boundary stress tensor $T_{\mu\nu}$ and Lagrangian $\mathcal{L}$, all to second order in boundary derivatives. These are the main results of this paper.

\subsection{The metric and dilaton}
This subsection contains our results for the metric and dilaton field. These expressions were obtained using a Weyl covariant form of the procedure outlined in detail in \cite{B1} (see also \cite{B2,B3} for similar calculations).
\nopagebreak
\begin{equation}\label{metric}
\begin{aligned}
ds^2&=-2 u_\mu dx^\mu \left[ dr + \left(r A_\nu + \frac{r^2 f(br)}{2} u_\nu\right) dx^\nu \right] +\left[r^2 P_{\mu\nu}+ 2(br)^2 F(br) \frac{1}{b}  \sigma_{\mu\nu}\right]dx^\mu dx^\nu \\
&+\left[\frac{1}{d-2}\mathcal{D}_\lambda \sigma^{\lambda}{}_{(\mu}u_{\nu)} +\frac{2L(br)}{(br)^{d-2}}u_{(\mu}P_{\nu )}^\lambda \mathcal{D}_\alpha \sigma^\alpha_\lambda - \frac{1}{d-2}\mathcal{D}_\lambda \omega^{\lambda}{}_{(\mu}u_{\nu)}\right]dx^\mu dx^\nu \\
&-\frac{2}{(br)^{d-2}} \left[ \frac{(br)^{d-2}}{2(d-2)} + L(br)\right]u^\alpha\mathcal{D}_\alpha \phi u_{( \mu} P_{\nu )} ^\lambda \mathcal{D}_\lambda \phi dx^\mu dx ^\nu\\
&-\left[\frac{1}{2(br)^d}\omega_{\alpha \beta}\omega^{\alpha \beta} +\frac{K_2(br)}{(br)^{d-2}}\frac{\sigma_{\alpha \beta}\sigma^{\alpha \beta}}{(d-1)} +\frac{\mathcal{R}}{(d-1)(d-2)}    \right] u_{\mu}u_{\nu}dx^\mu dx^\nu\\ 
&+ \frac{1}{2(d-2)(d-1)}P^{\alpha\beta} \mathcal{D}_\alpha \phi \mathcal{D}_\beta \phi u_{\mu}u_{\nu}dx^\mu dx^\nu\\ 
&+\frac{1}{(br)^{d-2}} \left[\frac{(2d-3)(br)^{d-2}}{2(d-1)(d-2)} +K_3(br) \right] u^\alpha u^\beta \mathcal{D}_\alpha \phi \mathcal{D}_\beta \phi u_{\mu}u_{\nu}dx^\mu dx^\nu\\
&+2(br)^2 \left[\left(F^2(br)-H_1(br)\right) \sigma_{\mu}{}^{\lambda}\sigma_{\lambda \nu}   +\left(H_2(br)-H_1(br)\right)u^{\lambda}\mathcal{D}_{\lambda}\sigma_{\mu \nu}\right]dx^\mu dx^\nu \\
&+2(br)^{2} \left[H_2(br)\left(\omega_{\mu}{}^{\lambda}\sigma_{\lambda \nu}+\omega_\nu{}^\lambda \sigma_{\lambda\mu}\right) -H_1(br) C_{\mu\alpha\nu\beta}u^\alpha u^\beta  \right]  dx^{\mu} dx^{\nu} -\omega_{\mu}{}^{\lambda}\omega_{\lambda\nu} dx^\mu dx^\nu\\
&+\frac{(br)^2} {(d-2)}H_1(br)\left[P^\alpha_\mu P^\beta_\nu\mathcal{D}_\alpha \phi \mathcal{D}_\beta \phi - \frac{1}{d-1} P_{\mu\nu}P^{\alpha\beta}\mathcal{D}_\alpha \phi \mathcal{D}_\beta \phi\right] dx^\mu dx^\nu \\
&+2(br)^2 \left[H_1(br)-K_1(br)\right]\frac{\sigma_{\alpha \beta}\sigma^{\alpha \beta}}{d-1}  P_{\mu\nu} dx^\mu dx^\nu -\frac{(br)^2}{d-1} K_1 (br) u^\alpha u^\beta \mathcal{D}_\alpha \phi \mathcal{D}_\beta \phi P_{\mu\nu} dx^\mu dx^\nu.
\end{aligned}
\end{equation}
\nopagebreak
The various functions which appeared in the metric above are defined as follows:
\begin{equation*}\label{metricfns:eq}
\begin{split}
F(br)&\equiv \int_{br}^{\infty}\frac{y^{d-1}-1}{y(y^{d}-1)}dy \\
&\approx  \frac{1}{br} -\frac{1}{d(br)^d}+ \frac{1}{(d+1)(br)^{d+1}}+\frac{\#}{(br)^{2d}}+\ldots\\
\end{split}
\end{equation*}
\begin{equation*}
\begin{split}
H_1(br)&\equiv \int_{br}^{\infty}\frac{y^{d-2}-1}{y(y^{d}-1)}dy \\
&\approx \frac{1}{2(br)^2}-\frac{1}{d(br)^d}+ \frac{1}{(d+2)(br)^{d+2}}+\frac{\#}{(br)^{2d}}+\ldots\\
\end{split}
\end{equation*}
\begin{equation*}
\begin{split}
H_2(br)&\equiv \int_{br}^{\infty}\frac{d\xi}{\xi(\xi^d-1)}
\int_{1}^{\xi}y^{d-3}dy \left[1+(d-1)y F(y) +2 y^{2} F'(y) \right]\\
&=\frac{1}{2} F(br)^2-\int_{br}^{\infty}\frac{d\xi}{\xi(\xi^d-1)}
\int_{1}^{\xi}\frac{y^{d-2}-1}{y(y^{d}-1)}dy\\
&\approx \frac{1}{2(br)^2}-\frac{1}{d(br)^d}\int_{1}^{\infty}\frac{y^{d-2}-1}{y(y^{d}-1)}dy\\
&\qquad -\frac{1}{d(br)^{d+1}}+\frac{3 d + 5}{2(d+1)(d+2)(br)^{d+2}}+\frac{\#}{(br)^{2d}}+\ldots \\
\end{split}
\end{equation*}
\begin{equation*}
\begin{split}
K_1(br) &\equiv \int_{br}^{\infty}\frac{d\xi}{\xi^2}\int_{\xi}^{\infty}dy\ y^2 F'(y)^2 \\
&\approx \frac{1}{2(br)^2}-\frac{2}{d(d+1)(br)^{d+1}}+\frac{2}{(d+1)(d+2)(br)^{d+2}}+\frac{\#}{(br)^{2d}}+\ldots\\
\end{split}
\end{equation*}
\begin{equation*}
\begin{split}
K_2(br) &\equiv \int_{br}^{\infty}\frac{d\xi}{\xi^2}\left[1-\xi(\xi-1)F'(\xi) -2(d-1)\xi^{d-1} \right.\\
&\left. \quad +\left(2(d-1)\xi^d-(d-2)\right)\int_{\xi}^{\infty}dy\ y^2 F'(y)^2 \right]\\
&\approx -\frac{(d-3)(d-1)}{2(d+1)(br)^2}+\frac{2(d-2)}{d(br)}+\frac{1}{d(2d-1)(br)^{d}}+\frac{\#}{(br)^{d+2}}+\ldots\\
\end{split}
\end{equation*}
\begin{equation*}
\begin{split}
K_3(br) &\equiv \frac{d-2}{2(d-1)} K_1 (br) - \frac{1}{d-1} F(br) + \frac{1}{2(d-1)} H_1 (br) \\
& \quad + \int_{br}^{\infty} d\xi \left( \xi^{d-3} - \xi^{d-2}\int_{\xi}^{\infty}dy\ y^2 F'(y)^2 \right)  \\
&\approx \frac{d-2}{d(d-1)(br)}+\frac{d-3}{4(d+1)(br)^2}+\frac{1}{2d(d-1)(2d-1)(br)^{d}}+\frac{\#}{(br)^{d+1}}+\ldots 
\end{split}
\end{equation*}
\begin{equation*}
\begin{split}
L(br) &\equiv \int_{br}^\infty\xi^{d-1}d\xi\int_{\xi}^\infty dy\ \frac{y-1}{y^3(y^d
-1)} \\
&\approx -\frac{1}{d(d+2)(br)^2}+\frac{1}{(d+1)(br)}\\
&\qquad-\frac{1}{(d+1)(2d+1)(br)^{d+1}}-\frac{1}{2(d+1)(d+2)(br)^{d+2}} +\frac{\#}{(br)^{2d}}+\ldots \\
\end{split}
\end{equation*}
Their asymptotic forms at large $r$ are also provided; these will be required to calculate the corresponding boundary stress tensor and Lagrangian.

Note that, as concluded in our discussion of manifest Weyl covariance in section \ref{3}, the bulk metric can be written in the following Weyl covariant form:
\begin{equation*}
ds^2 = -2 u_\mu dx^\mu(dr+\mathcal{V}_\nu dx^\nu)+\mathcal{G}_{\mu\nu}dx^\mu dx^\nu,
\end{equation*}
where the functions $\mathcal{V}_\mu$ and $\mathcal{G}_{\mu\nu}$ are given by:
\begin{equation*}
\begin{split}
\mathcal{V}_\mu &=  r A_\mu + \frac{r^2 f(br)}{2}u_\mu  \\
&-\frac{1}{2(d-2)}\mathcal{D}_\lambda \sigma^{\lambda}{}_{\mu} -\frac{L(br)}{(br)^{d-2}}P_{\mu }^\lambda \mathcal{D}_\alpha \sigma^\alpha_\lambda + \frac{1}{2(d-2)}\mathcal{D}_\lambda \omega^{\lambda}{}_{\mu}\\
&+\frac{1}{(br)^{d-2}} \left[ \frac{(br)^{d-2}}{2(d-2)} + L(br)\right]u^\alpha\mathcal{D}_\alpha \phi P_{\mu } ^\lambda \mathcal{D}_\lambda \phi \\
&+\left[\frac{1}{4(br)^d}\omega_{\alpha \beta}\omega^{\alpha \beta} +\frac{K_2(br)}{2(br)^{d-2}}\frac{\sigma_{\alpha \beta}\sigma^{\alpha \beta}}{(d-1)} +\frac{\mathcal{R}}{2(d-1)(d-2)}    \right] u_{\mu}\\ 
&- \frac{1}{4(d-2)(d-1)}P^{\alpha\beta} \mathcal{D}_\alpha \phi \mathcal{D}_\beta \phi u_{\mu}\\ 
&-\frac{1}{2(br)^{d-2}} \left[\frac{(2d-3)(br)^{d-2}}{2(d-1)(d-2)} +K_3(br) \right] u^\alpha u^\beta \mathcal{D}_\alpha \phi \mathcal{D}_\beta \phi u_{\mu}  +\cdots  ,
\end{split}
\end{equation*}
\begin{equation}
\begin{split}
\mathcal{G}_{\mu\nu}&= r^2 P_{\mu\nu}-\omega_{\mu}{}^{\lambda}\omega_{\lambda\nu}\\
&+2(br)^2 F(br)\left[\frac{1}{b}  \sigma_{\mu\nu} +  F(br)\sigma_{\mu}{}^{\lambda}\sigma_{\lambda \nu}\right]-2(br)^2 K_1(br)\frac{\sigma_{\alpha \beta}\sigma^{\alpha \beta}}{d-1}P_{\mu\nu}\\
&-2(br)^2 H_1(br)\left[u^{\lambda}\mathcal{D}_{\lambda}\sigma_{\mu \nu}+\sigma_{\mu}{}^{\lambda}\sigma_{\lambda \nu} -\frac{\sigma_{\alpha \beta}\sigma^{\alpha \beta}}{d-1}P_{\mu \nu} + C_{\mu\alpha\nu\beta}u^\alpha u^\beta \right]\\
&+2(br)^{2} H_2(br)\left[u^{\lambda}\mathcal{D}_{\lambda}\sigma_{\mu \nu}+\omega_{\mu}{}^{\lambda}\sigma_{\lambda \nu}+\omega_\nu{}^\lambda \sigma_{\mu\lambda}\right] \\
&+\frac{(br)^2} {(d-2)}H_1(br)\left[P^\alpha_\mu P^\beta_\nu\mathcal{D}_\alpha \phi \mathcal{D}_\beta \phi - \frac{1}{d-1} P_{\mu\nu}P^{\alpha\beta}\mathcal{D}_\alpha \phi \mathcal{D}_\beta \phi\right]  \\
& -\frac{(br)^2}{d-1} K_1 (br) u^\alpha u^\beta \mathcal{D}_\alpha \phi \mathcal{D}_\beta \phi P_{\mu\nu} +\cdots .
\end{split}
\end{equation}

And finally, we present our result for the dilaton to second order:
\begin{equation}
\Phi=\phi + b u^\mu \mathcal{D}_\mu \phi F(br) + \frac{b^2}{d-2}  H_1(br) \mathcal{D}^2 \phi +b^2 H_2 (br) u^\mu u^\nu \mathcal{D}_\mu \phi \mathcal{D}_\nu \phi.
\end{equation}

The expressions listed here have been compared against existing results in the literature \cite{B1, B2,B3,Haack} and wherever there is an overlap, we find complete agreement. Certain papers have performed similar calculations but in a different gauge \cite{B1,B2}; and so, when comparing our results with these papers, we have utilised the appropriate gauge transformation.

\subsection{Stress tensor and Lagrangian of the dual fluid}

We now calculate the boundary stress tensor and Lagrangian using the standard AdS/CFT formulae:
\begin{equation}
\begin{split}
16\pi G_{d+1} T^\mu_\nu &= \lim_{r\rightarrow\infty} r^d\left(2(K_{\alpha\beta}h^{\alpha\beta} \delta^\mu_\nu - K^\mu_\nu)\right.\\
&\left.+\bar{\mathfrak{G}}^\mu_\nu-\frac{d(d-1)}{2} \delta^\mu_\nu  -\frac{1}{d-2}\left(\bar{\nabla}^\mu \Phi\bar{\nabla}_\nu\Phi-\frac{\delta^\mu_\nu}{2} (\bar{\nabla} \Phi)^2  \right) \right),\\
16\pi G_{d+1} e^{-\phi}\ {\cal L} &= - \lim_{r\rightarrow\infty} r^d\left(\partial_n\Phi+\frac{1}{d-2}\bar{\nabla}^2\Phi\right).
\end{split}
\end{equation}
Here, $h_{\mu\nu}$ is the induced metric on the constant $r$ hypersurface; from this, we obtain the covariant derivative $\bar{\nabla}$ and the corresponding Einstein tensor $\bar{\mathfrak{G}}^\mu_\nu$. We define $n^A$ to be the outward pointing unit normal of the constant $r$ hypersurface; the extrinsic curvature of the constant $r$ hypersurface is then defined by the Lie derivative of the induced metric,  $K_{\mu\nu}\equiv \frac{1}{2} \mathfrak{L}_n h_{\mu\nu} $, and $\partial_n$ is the partial derivative along $n^A$. In the formulae above, all the indices are raised using the induced metric.

We find that the boundary stress tensor is given by:
\begin{equation}
\begin{split}
16\pi G_{d+1} T_{\mu\nu} &= b^{-d}\left(g_{\mu\nu}+d u_\mu u_\nu \right)-2 b^{1-d} \sigma_{\mu\nu}\\
&-2 b^{2-d} \tau_\omega \left[u^{\lambda}\mathcal{D}_{\lambda}\sigma_{\mu \nu}+\omega_{\mu}{}^{\lambda}\sigma_{\lambda \nu}+\omega_\nu{}^\lambda \sigma_{\mu\lambda} \right]\\
&+2b^{2-d}\left[u^{\lambda}\mathcal{D}_{\lambda}\sigma_{\mu \nu}+\sigma_{\mu}{}^{\lambda}\sigma_{\lambda \nu} -\frac{\sigma_{\alpha \beta}\sigma^{\alpha \beta}}{d-1}P_{\mu \nu}+ C_{\mu\alpha\nu\beta}u^\alpha u^\beta \right]\\
&-\frac{1}{d-2} b^{2-d}\left[ P^\alpha_\mu P^\beta_\nu\mathcal{D}_\alpha \phi \mathcal{D}_\beta \phi - \frac{1}{d-1} P_{\mu\nu}P^{\alpha\beta}\mathcal{D}_\alpha \phi \mathcal{D}_\beta \phi\right]
\end{split}
\end{equation}
with
\begin{equation}
b=\frac{d}{4\pi T}\qquad \text{and} \qquad \tau_{\omega} =   \int_{1}^{\infty}\frac{y^{d-2}-1}{y(y^{d}-1)}dy.
\end{equation}
We further obtain the following expression for the Lagrangian:
\begin{equation}
16 \pi G_{d+1} e^{-\phi}\mathcal{L} = -b^{1-d} u^\mu \mathcal{D}_\mu \phi - \frac{1}{d-2} b^{2-d} \mathcal{D}^2 \phi - b^{2-d} \tau_\omega u^\mu u^\nu \mathcal{D}_\mu \phi \mathcal{D}_\nu \phi.
\end{equation}
And again, these results are all consistent with the existing literature \cite{B1,B2,B3,Haack}.

\section{Discussion}\label{5}
In this paper, we have constructed asymptotically locally AdS$_{d+1}$ bulk spacetimes with a slowly varying dilaton field which are dual, under the AdS/CFT correspondence, to forced fluid flows on the weakly curved boundary metric. These forced fluid flows satisfy the conformal relativistic Navier-Stokes equations with a dilaton-dependent forcing term. We have also obtained the form of the dual stress tensor and Lagrangian, all to second order in the boundary derivative expansion. Our results further generalise previous work on the fluid/gravity correspondence \cite{B1,B2,B3,Haack}; in particular, we have generalised the results of \cite{B3} to arbitrary spacetime dimensions. 

There are several interesting applications of our paper which merit further consideration. It would be useful to study in detail holographic models of novel forced fluid flows. One avenue which is worth exploring would be to consider holographic duals of forced fluid flows which exhibit turbulence.
A holographic model of turbulence would certainly offer a new perspective on this poorly understood phenomenon, and hopefully new and fruitful insights could then be derived from this. By carefully choosing the form of the forcing term (which is fixed by our choice of $\phi(x)$), it could be possible to stir the boundary fluid into turbulent configurations\footnote{Another possibility of realising turbulence which is potentially more straightforward would be to consider a boundary spacetime which consists of small time-dependent fluctuations away from flat space. These linearised time-dependent fluctuations can effectively act as a forcing term for a fluid on a flat background metric; this is discussed in more detail in the introduction of \cite{B2}.}. A noteworthy point that should be raised here is that unforced fluid flows can exhibit turbulence as well; however, these turbulent phases will be transient. And, in fact, a holographic model of transient turbulence has already been constructed \cite{Liu3}. The key advantage of considering holographic models of forced fluid flows, on the other hand, is the possibility of realising holographic models of \textit{steady state} turbulence; such configurations can only exist with a forcing term as the fluid would otherwise eventually settle down into a non-dissipative configuration.

Also, observe that the expressions that we have obtained are valid for arbitrary spacetime dimensions; this is particularly relevant for the study of turbulence. It is a well-known fact that turbulent phases for relativistic fluids in $2+1$ dimensions display remarkably different behaviour to relativistic fluids in higher dimensions. In $2+1$ dimensions, relativistic fluids display an `inverse energy cascade'; energy cascades from short to long wavelengths \cite{Myers}. This is in sharp contrast to the standard cascade observed in higher dimensions which is from long to short wavelengths. The results of this paper could in principle be used to construct holographic models of turbulence in different dimensions which would then shed light on the source of the discrepancy between the nature of the energy cascades in two spatial dimensions and greater. Such holographic models would also be of interest purely from a gravitational perspective. The construction of such models would suggest that AdS$_4$ displays qualitatively different instabilities to AdS$_{d+1}$ for $d>3$. Further, these models may have interesting connections to the weakly turbulent instability of AdS discovered in \cite{AdS}. We will be investigating several different approaches to holographic turbulence in future work.

\acknowledgments
It is a pleasure to thank Malcolm Perry for proofreading this paper. We further thank Rajesh Gopakumar, Veronika Hubeny, Shiraz Minwalla, and Mukund Rangamani for comments on a draft of this paper. Ashok is supported by the Cambridge Commonwealth, European, and International Trust, DAMTP, and Trinity College, Cambridge. 

\appendix

\section{First order calculation}
In these appendices, we present further details of our calculations. As explained in section \ref{2}, the equations involved can be classified either as constraint equations which impose relations between the fluid dynamical variables at one order lower, or as dynamical equations which enable us to solve for the metric at the order at which we are working. In this first appendix, we present these equations at first order; the following appendix contains the analogous equations at second order.

\subsection{Constraint equations}
The constraint equations are obtained by contracting the Einstein equations (\ref{Einstein-dilaton}) with $n^A$, the normal to the constant $r$ hypersurface: $E^\Phi_{AB} n^B=0$. The boundary components of this equation give us the following relation: 
\begin{equation} \label{constraint}
\partial_\mu b = \mathcal{A}_\mu b.
\end{equation}
It is not difficult to show that this relation is equivalent to the equations of forced fluid dynamics $ \nabla_\mu T^{\mu\nu}=e^{-\phi} \mathcal{L} \nabla^\nu \phi$. This explicitly confirms our expectation that our perturbative procedure constructs bulk spacetimes dual to solutions of forced fluid dynamics. At first order, the relevant terms in this equation stem from the zeroth order contributions to the stress tensor $T^{\mu\nu}$ and Lagrangian $\mathcal{L}$; and since the Lagrangian only consists of first and higher orders terms, the equations at this order are the same as the unforced case. 

\subsection{Dynamical equations and source terms}
We now present the dynamical equations at first order. Solving these equations subject to regularity away from $r>0$ and normalisability on the boundary will allow us to obtain the first order contributions to the bulk metric (\ref{metric}). In the equations that follow (in both this appendix and the subsequent one), we will use $\mathcal{V}_\mu^{(i)}$ and $\mathcal{G}_{\mu\nu}^{(i)}$ to refer to the $i^{\text{th}}$ order contributions to $\mathcal{V}_\mu$ and $\mathcal{G}_{\mu\nu}$ respectively.

It turns out that at first order, there is only one equation with nonzero source terms, and this is given by the transverse traceless part of $E^\Phi_{\mu\nu}=0$:
\begin{equation}
\begin{aligned}
-\frac{(br)^2 -(br)^{2-d}}{2} \mathcal{G}^{(1)\prime\prime}_{\mu\nu} - \frac{(br)}{2}(d-3) \mathcal{G}^{(1)\prime}_{\mu\nu} &- \frac{3(br)^{1-d}}{2} \mathcal{G}^{(1)\prime}_{\mu\nu} + (d-2+2(br)^{-d})\mathcal{G}^{(1)}_{\mu\nu} \\
&= (d-1) (br) \sigma_{\mu\nu}.
\end{aligned}
\end{equation}
Here, we have used $\tilde{\mathcal{G}}_{\mu\nu} \equiv \mathcal{G}_{\mu\nu} - \frac{1}{d-1}\mathcal{G}^\alpha_\alpha P_{\mu\nu}$; further, the ${}^\prime$ refers to a derivative with respect to $br$.

\section{Second order calculation}

\subsection{Constraint equations}
If we now consider the boundary components of $E^\Phi_{AB} n^B=0$ at second order, we find that the relation (\ref{constraint}) acquires a second order correction:
\begin{equation}
\partial_\mu b = \mathcal{A}_\mu b + 2 b^2 \left[ \frac{\sigma_{\alpha\beta} \sigma^{\alpha\beta}}{d-1}u_\mu - \frac{\mathcal{D}_\lambda \sigma^\lambda {}_\mu}{d} \right] + \frac{b^2}{d}\left[\frac{u^\alpha\mathcal{D}_\alpha \phi u^\beta\mathcal{D}_\beta \phi}{d-1}u_\mu + u^\alpha\mathcal{D}_\alpha \phi P_{\mu\lambda} \mathcal{D}^\lambda \phi\right].
\end{equation}
This can again be shown to be equivalent to $ \nabla_\mu T^{\mu\nu}=e^{-\phi} \mathcal{L} \nabla^\nu \phi$; the additional second order terms arise from the first order contributions to $T^{\mu\nu}$ and $\mathcal{L}$. 

\subsection{Dynamical equations and source terms}
At second order, there are several dynamical equations which need to be considered. To obtain the trace part of $\mathcal{G}_{\mu\nu}^{(2)}$, we use the equation $E^\Phi_{rr}=0$:
\begin{equation}
\begin{aligned}
-\frac{(br)^2}{2} \mathcal{G}^{(2)\prime\prime}{}^\mu_\mu+(br)\mathcal{G}^{(2)\prime}{}^\mu_\mu -\mathcal{G}^{(2)}{}^\mu_\mu &=\left[-4(br)^3F(br)F^\prime(br) -(br)^4F^{\prime 2}(br) \right. \\
&\left.- 2(br)^4F^{\prime\prime}(br)F(br)\right] \sigma_{\mu\nu}\sigma^{\mu\nu} - \omega_{\mu\nu}\omega^{\mu\nu}\\
&+\frac{(br)^4}{2}u^\mu\mathcal{D}_\mu \phi u^\nu \mathcal{D}_\nu \phi F^{\prime 2} (br),
\end{aligned}
\end{equation}
while the traceless part $\tilde{\mathcal{G}}_{\mu\nu}^{(2)}$ is obtained by considering the transverse traceless part of $E^\Phi_{\mu\nu}=0$:
\begin{equation}
\begin{aligned}
&\left(-\frac{1}{2}(br)^2+\frac{1}{2}(br)^{2-d}\right)\tilde{\mathcal{G}}^{(2)\prime\prime}_{\mu\nu}+\left(\frac{(3-d)}{2}(br)-\frac{3}{2}(br)^{1-d}\right)\tilde{\mathcal{G}}^{(2)\prime}_{\mu\nu}+\left((d-2)+2(br)^{-d}\right)\tilde{\mathcal{G}}^{(2)}_{\mu\nu}\\
&=\left[-(d-4)+2(d-1)(br)F(br)+4(br)^2F^\prime(br)+\frac{2(1-(br)^{2-d})}{(br)^d-1}\right]\left(\sigma_{\mu}{}^{\lambda}\sigma_{\lambda \nu} -\frac{\sigma_{\alpha \beta}\sigma^{\alpha \beta}}{d-1}P_{\mu \nu}\right)\\
&+\left[-(d-3)+(d-1)(br)F(br)+2(br)^2F^\prime(br)\right]u^\lambda\mathcal{D}_\lambda\sigma_{\mu\nu}\\
&+\left[1+(d-1)(br)F(br)+2(br)^2F^\prime(br)\right]\left(\omega_{\mu}{}^{\lambda}\sigma_{\lambda \nu}+\omega_\nu{}^\lambda \sigma_{\mu\lambda}\right)\\
&+\left[-(d-2)-2(br)^{-d}\right]\omega_\mu{}^\lambda\omega_{\lambda\nu} +(d-2)C_{\mu\alpha\nu\beta}u^\alpha u^\beta\\
&-\frac{1}{2}\left(P^\alpha_\mu P^\beta_\nu\mathcal{D}_\alpha \phi \mathcal{D}_\beta \phi - \frac{1}{d-1} P_{\mu\nu}P^{\alpha\beta}\mathcal{D}_\alpha \phi \mathcal{D}_\beta \phi\right) .
\end{aligned}
\end{equation}
Further, the transverse part of the equation $E^\Phi_{r\mu}=0$ gives us $P_\mu^\nu\mathcal{V}^{(2)}_\nu$:
\begin{equation}
\begin{aligned}
&\frac{1}{2}(br)^2P_\mu^\lambda\mathcal{V}_\lambda^{(2)\prime\prime} +\frac{d-3}{2}(br)P_\mu^\lambda\mathcal{V}_\lambda^{(2)\prime} - (d-2)P_\mu^\lambda\mathcal{V}_\lambda^{(2)}\\
&=\frac{(br)^2}{2}F^{\prime}(br)u^\alpha\mathcal{D}_\alpha \phi P_{\mu } ^\lambda \mathcal{D}_\lambda \phi
-\frac{(br)^2}{2}F^{\prime}(br)P_\mu^\lambda\mathcal{D}_\alpha \sigma^{\alpha}_\lambda -\frac{1}{2(br)}P_\mu^\lambda\mathcal{D}_\alpha \omega^{\alpha}{}_\lambda
,
\end{aligned}
\end{equation}
while $u^\alpha\mathcal{V}_\alpha^{(2)}$ is obtained from the trace of the transverse part of $E^\Phi_{\mu\nu}=0$:
\begin{equation}
\begin{aligned}
&2(br)u^\lambda\mathcal{V}_\lambda^{(2)\prime}+(d-2)2u^\lambda\mathcal{V}_\lambda^{(2)}
-\frac{(br)^2-(br)^{2-d}}{2(d-1)}\mathcal{G}^{(2)\prime\prime}{}^\alpha_\alpha \\
&+\left(\frac{2-d}{d-1}(br)+\frac{d-4}{2(d-1)}(br)^{1-d}\right)\mathcal{G}^{(2)\prime}{}^\alpha_\alpha 
+\left(\frac{2d-3}{d-1}-\frac{d-3}{d-1}(br)^{-d}\right)\mathcal{G}^{(2)}{}^\alpha_\alpha\\
&=-u^\alpha\mathcal{D}_\lambda\sigma^\lambda_\alpha+u^\alpha\mathcal{D}_\lambda\omega^\lambda{}_\alpha -\frac{\mathcal{R}}{d-1} +\frac{1}{2(d-1)}P^{\alpha\beta}\mathcal{D}_\alpha \phi \mathcal{D}_\beta \phi \\
&+\left[\frac{2(d-2)}{d-1}(br)^{-d}-\frac{2d-3}{d-1}\right]\omega_{\alpha\beta}\omega^{\alpha\beta}\\
& +F(br)\left[\frac{-4d}{d-1}(br)^3F^\prime(br) + \frac{2d}{d-1}(br)^{3-d}F^\prime (br) -\frac{2}{d-1}\left((br)^4-(br)^{4-d}\right)F^{\prime\prime}(br)\right]\sigma_{\alpha\beta}\sigma^{\alpha\beta}\\
&+\left[\frac{-1}{d-1}F^{\prime 2}(br)\left((br)^4-(br)^{4-d}\right)+\frac{1}{d-1}(br)^{1-d}-\frac{1}{d-1}(br)^{2-d}(br-1)F^\prime (br) -2\right]\sigma_{\alpha\beta}\sigma^{\alpha\beta}\\
&+\left[\frac{-1}{2(d-1)}-\frac{1}{d-1}(br)^{3-d}F^\prime(br) - \frac{1}{2(d-1)}(br)^{2-d}\frac{(br)^{d-2}-1}{(br)^d-1}\right]u^\alpha u^\beta \mathcal{D}_\alpha \phi \mathcal{D}_\beta \phi \\
&+\left[\frac{1}{2(d-1)}\left((br)^4-(br)^{4-d}\right)F^{\prime 2}(br)\right]u^\alpha u^\beta \mathcal{D}_\alpha \phi \mathcal{D}_\beta \phi.
\end{aligned}
\end{equation}

\end{document}